\documentclass[12pt,a4paper]{article}
\usepackage{latexsym,amssymb,amsmath,amsthm}
\usepackage{math}

\usepackage{fourier}
\usepackage[usenames]{color}
\usepackage{color}
\definecolor{black}{rgb}{0.0,0.0,0.0}
\definecolor{white}{rgb}{1.0,1.0,1.0}
\definecolor{vertfonce}{rgb}{0.20, 0.46, 0.25}
\definecolor{rougefonce}{rgb}{0.64, 0.09, 0.20}
\definecolor{bleu}{rgb}{0.0,0.1,1.0}
\usepackage[colorlinks,citecolor=blue,urlcolor=blue]{hyperref}
\usepackage{tikz}
\usetikzlibrary{arrows,positioning}

\begin{document}
\begin{center}
{\Large \textsf{\textbf{On the entropy of the mean field spin glass model
                              }}}\\
  \vskip1.5cm
Flora \textsc{Koukiou}

{\small
Laboratoire de physique th\'eorique et mod\'elisation
(CNRS UMR 8089)\\
Universit\'e de Cergy-Pontoise
F-95302 Cergy-Pontoise\\
\texttt{flora.koukiou@u-cergy.fr}\\
\vskip5mm
1st July 2012}
\end{center}

\noindent
{
\textbf{Abstract:} From the study of a functional equation relating the Gibbs measures at two different tempratures we prove that the specific entropy of the Gibbs measure of the Sherrington-Kirkpatrick Spin Glass Model vanishes at the inverse temperature $\beta=4 \log2$.
}
\vskip0.5cm
\noindent

\section{Introduction and main results}

Over the last decade, mean field models of spin glasses
have motivated  increasingly many
studies by   physicists and mathematicians \cite{ALR,G,GT, K1,K2,
P, PS, T1}.
The existence of infinite volume limit of thermodynamic quantities is now
rigorously  established thanks to the development of numerous remarkable 
 analytical techniques.
For the Sherrington-Kirkpatrick model, the first major results of   
Guerra and Toninelli \cite{GT} 
on   existence and uniqueness  of the free energy, 
 are generalized by 
Aizenman, Sims and Starr \cite{ASS} in a  scheme giving variational
upper bounds on the free energy.  Talagrand \cite{T1}, under some conditions on the overlap function, contributed to the entirely rigorous account  of the  original formulae proposed 
by Parisi \cite{P}.

An interesting question, related to the
  behaviour of  Gibbs measures, is the study of their specific entropy.
Despite the numerous developments achieved lately on this model, the study of the properties of the entropy is still missing in the literature.
The specific entropy  decreases with the temperature and   the high temperature entropy  can easily be estimated.  By lowering the temperature the entropy should eventually vanish and an
early  result, given in \cite{ALR}, corroborates the idea  that the entropy does not vanish very fast.
In this note we estimate the value of the (low) temperature at which the mean entropy  of the Gibbs measure vanishes.

The approach we use here is totally self-contained. From the low-temperature results, we need solely the existence of the thermodynamic limit of the quenched specific free energy and its self-averaging property.

We first recall some basic definitions.
Suppose that a finite set of $n$ sites is given. With each site we associate the one-spin space $\Sigma:=\{1, -1\}$. The natural configuration space is then the  product space  $\Sigma_n=\{-1,1\}^n=\Sigma^n$, with $\card\Sigma_n =2^n$ equipped with the uniform probability measure $\nu_n$.
For each $\sigma\in\Sigma_n$,
the finite volume Hamiltonian of the model 
is given by the following real-valued function on $\Sigma_n$
\[H_n(\sigma)=-\frac{1}{\sqrt{n}}
\sum_{1\leq i<j\leq n}J_{ij}\sigma_i\sigma_j,\]
where the family of couplings $J=(J_{ij})_{1\leq i<j\leq n}$ are 
independent  centred Gaussian random variables of variance $1$.

At the  inverse temperature $\beta=\frac{1}{T}>0$, 
the disorder dependent  partition function $Z_n(\beta,J)$, 
  is given by the sum of the Boltzmann factors
  \[Z_n(\beta, J)=\sum_{\sigma}e ^{-\beta H_n(\sigma,J)}.\]
Moreover,  if
$E_J$ denotes the expectation with respect to the randomness $J$,
 it is very simple to show that  $E_JZ_n(\beta,J)=2^n e^{\frac{\beta^2}{4}(n-1)}$.  
 
 When the randomness $J$ is fixed, the corresponding conditional Gibbs probability measure is denoted  
by $\mu_{n,\beta}(\sigma|J)$ and given by:
\[\mu_{n,\beta}(\sigma|J)=
\frac{e^{-\beta H_n(\sigma,J)}}{Z_n(\beta, J)}.\]
The entropy of $\mu_{n,\beta}$, is  defined as usual by
$S(\mu_{n,\beta,J})=-\sum_\sigma\mu_{n,\beta}(\sigma|J)
 \log  \mu_{n,\beta}(\sigma|J)$.
 
The  real functions  
\[f_n(\beta)=\frac{1}{n}E_J\log Z_n(\beta,J)\] and
\[\overline{f}_n(\beta)=\frac{1}{n}\log E_J Z_n(\beta,J),\]
define  the quenched average of the specific free energy and 
 the annealed  specific free energy respectively. 
 The ground state energy density $-\epsilon_n(J)$ is given by
\[-\epsilon_n(J)=\frac{1}{n}\inf_{\sigma\in\Sigma_n} H_n(\sigma,J).\]

At the low temperature region ($\beta>1$),
the following two infinite volume limits
  \[\lim_{n\rightarrow\infty}f_n(\beta, J)=f_\infty(\beta),\] and, 
\[-\lim_{n\rightarrow\infty}\epsilon_n(J)=
\lim_{\beta\rightarrow\infty}\frac{f_\infty(\beta)}{\beta}=-\epsilon_0\] exist for almost all $J$ and are non random; this result has been rigorously  proved by Guerra and Toninelli \cite{GT}.

The main results of this note are stated in the following and proved in the next section.

\noindent
{\bf Proposition:}
{\it Almost surely, at the inverse temperature  $\beta_*=4 \log 2=2.77258\cdots$, the thermodynamic limit of the quenched free energy is given by
\[f_\infty(\beta_*)=\lim_{n\rightarrow\infty}\frac{1}{n}E_J \log Z_n(\beta_*,J)=f_\infty(1)+(\beta_*-1)\log 2=\frac{\beta_*^2}{4}+\frac{\beta_1^2}{4}=\beta_* \log 2+\frac{1}{4}.\]}

The Parisi formula provides with the expression of the free energy for the entire low temperature region in terms of a functional equation;   the 
pertinence of the precise calculation of the limit 
at a particular value of the temperature 
stems from its usefulness in determining the point where the entropy vanishes. This gives new insight to the behaviour of the model and is summarised in the following

\vskip0.6mm

\noindent
{\bf Theorem:}
{\it At the inverse temperature $\beta_*=4 \log 2=2.77258\cdots$, the specific entropy $s(\mu_{\beta_*})$ of the Gibbs measure vanishes almost surely:

\[s(\mu_{\beta_*}):=\lim_{n\rightarrow\infty}\frac{1}{n}
S(\mu_{n,\beta_*,J})=-\lim_{n\rightarrow\infty}\frac{1}{n}\sum_\sigma\mu_{n,\beta_*}(\sigma|J)
 \log  \mu_{n,\beta_*}(\sigma|J)=0.\]}
 
\Rk The formulation of the above  statement assumes  that the limit $\lim_{n\rightarrow\infty}\frac{1}{n}
S(\mu_{n,\beta_*,J})$ exists and is independent of $J$. This follows from general principles and can immediately be obtained from the existence and self-averaging of the low temperature specific free energy.

\cut{A directly  related result is the following corollary, which improves all the rigorous 
lower bounds for the ground state energy.\\

\noindent
{\bf Corollary:}
{\it The ground state energy 
 density  of the Sherrington-Kirkpatrick  
spin glass model is almost surely bounded  by
\[\epsilon_0\geq  = -0.7833\cdots.\]
}
}
\section{\bf Proof of the main results}
Notice first,  that for all  $\beta>0$,  the quenched limit   $f_\infty(\beta)$
exists and  is a convex function of $\beta$ \cite{GT}.
Let $
\beta_1\equiv 1$. From the high temperature results \cite{ALR}, 
we have, almost surely, that
\begin{eqnarray*}
f_\infty(\beta_1)
&=&
\lim_{n\rightarrow\infty}\frac{1}{n}
E_J\log Z_n(\beta_1,J)=
\lim_{n\rightarrow\infty}\frac{1}{n}\log E_JZ_n(\beta_1,J)\\
& = & \overline{f}_\infty(\beta_1)=
\log 2 +\frac{\beta_1^2}{4}\\
&=&\log 2+\frac{1}{4}.
\end{eqnarray*}
The following figure \ref{fig:beta}  illustrates the definition of the inverse temperature $\beta_*$;  the annealed free energy $\overline{f}_\infty(\beta)=
\log 2+\frac{\beta^2}{4}$ 
is plotted as a function of $\beta$ and the straight line is defined  by $\frac{\beta}{\beta_1} f_\infty(\beta_1)\equiv \beta  f_\infty(\beta_1)$.
The two graphs intersect at $\beta_1=1$ and $\beta_*=4\log 2=2,77258\cdots$.  One can now easily check  that, at  $\beta=\beta_*$,  the annealed free energy $\overline{f}_\infty(\beta_*)$ is simply related to $f_\infty(\beta_1)$  by  the following relationship
\[
\overline{f}_\infty(\beta_*)=\frac{\beta_*^2}{4}+\log 2=\frac{\beta_*}{\beta_1}(\frac{\beta_* \beta_1}{4}+
\frac{\beta_1}{\beta_*}\log 2)=\frac{\beta_*}{\beta_1}
(\log 2+\frac{1}{4})=\frac{\beta_*}{\beta_1} f_\infty(\beta_1).
\]

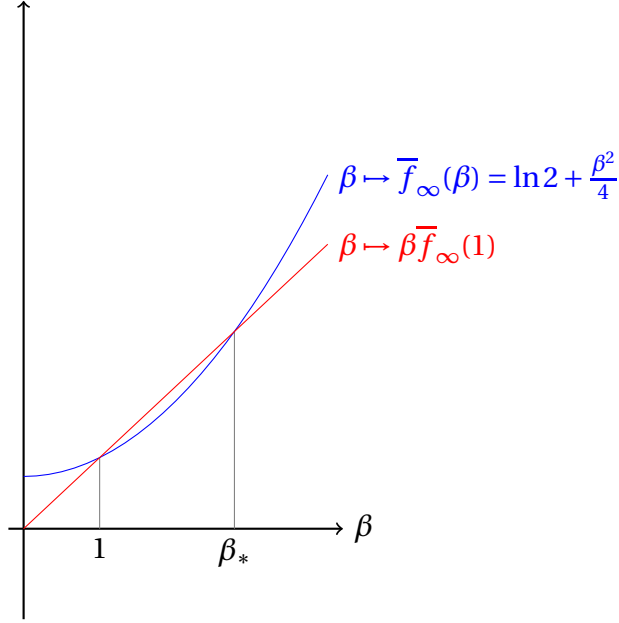
\begin{figure}[h]
\begin{center}
\begin{tikzpicture}[domain=0:4]
    \draw[->, thick,color=black] (-0.2,0) -- (4.2,0) node[right] {$\beta$};
    \draw[->, thick, color=black] (0,-1.2) -- (0,7);
   \draw[-, color=gray]    (1,0) -- (1, 0.943147) ;
   \draw[-] (1,0)--(1,0.001) node[below] {$1$};
   \draw[-] (2.772588,0)--(2.772588,0.001) node[below] {$\beta_*$};
    \draw[-, color=gray]    (2.772588,0) -- (2.772588,2.614959 ) ;
    \draw[color=blue]  plot[id=annealed] function{log(2)+0.25*(x**2)}
        node[right] {$\beta\mapsto\overline{f}_\infty(\beta) =\ln 2+ \frac{\beta^2}{4}$};
    \draw[color=red] plot[id=linear] function{x*(log(2)+0.25)} 
        node[right] {$\beta \mapsto \beta\overline{f}_\infty(1)$};
  \end{tikzpicture}
\end{center}
\caption{\label{fig:beta} 
The value  $\beta_*= 4\log 2$,  is given by the intersection of  the graph of the 
  annealed free energy $\overline{f}_\infty(\beta)$ 
  with the straight line  $\beta \overline{f}_\infty(1)$.}
\end{figure} 

We denote by $T$  the mapping   $T:  \mu_{n,\beta_1}(\sigma|J)\mapsto\mu_{n,\beta}(\sigma|J)$  defining, for all  $\beta>\beta_1$,
 the Gibbs probability  measure  
$\mu_{n,\beta}(\sigma|J)$  via the functional equation 
\[
\mu_{n,\beta}(\sigma|J):=\frac{\exp(-\beta H_n(\sigma, J))}{Z_n(\beta,J)}=\mu_{n,\beta_1}^{\beta/\beta_1}(\sigma|J)
\frac{Z_n^{\beta/\beta_1}(\beta_1,J)}{Z_n(\beta,J)}.
 \]
Notice that $\beta/\beta_1$ is  a non dimensional quantity. Moreover  the value $\beta_1$ fixes the temperature scale {\it i.e.} the temperature $\beta>\beta_1$ is expressed in units where $\beta_1\equiv 1$.

Since  $\mu_{n,\beta}$ is a probability on the configuration space, summing up over the configurations $\sigma$ and taking the thermodynamic limit, we have  indeed    
\[\lim_{n\rightarrow \infty}\frac{1}{n}\log\sum_\sigma\mu_{n,\beta}(\sigma|J)=
\lim_{n\rightarrow \infty}\frac{1}{n}
 \log \sum_\sigma\mu_{n,\beta_1}^{\beta/\beta_1}(\sigma|J)
+\alpha_\infty(\beta_1, \beta)=0,
\]
where  the limit  $\alpha_\infty(\beta_1, \beta)$
is given by

\begin{eqnarray*}
\alpha_\infty(\beta_1, \beta)
& = &
\lim_{n\rightarrow \infty}\frac{\beta}{\beta_1}\frac{1}{n}\log Z_n(\beta_1,J)-
\lim_{n\rightarrow \infty}\frac{1}{n}\log Z_n(\beta,J)\\
& = &
\lim_{n\rightarrow \infty}\frac{\beta}{\beta_1}\frac{1}{n}E_J\log Z_n(\beta_1,J)-
\lim_{n\rightarrow \infty}\frac{1}{n}E_J\log Z_n(\beta,J) \ \textrm{(due to the self-averaging)}\\
& = &
\frac{\beta}{\beta_1}f_\infty(\beta_1)-f_\infty(\beta).
\end{eqnarray*}
The existence, for all $\beta>\beta_1$, of  the limit $\alpha_\infty(\beta_1, \beta)$ follows immediately from the existence of the two
limits $f_\infty(\beta_1)$ and $f_\infty(\beta)$.
Now, 
by making use of the relation between the  limits 
$\overline{f}_\infty(\beta_*)$ and $f_\infty(\beta_1)$, one can check, that for  $\beta=\beta_*$, the limit  $\alpha_\infty(\beta_1, \beta_*)$ gives the deviation of the  free energy $f_\infty(\beta_*)$ from its mean value :
\[
a_\infty
: = \alpha_\infty(\beta_1, \beta_*)=\frac{\beta_*}{\beta_1}f_\infty(\beta_1)-f_\infty(\beta_*)
=\overline{f}_\infty(\beta_*)-f_\infty(\beta_*).
\]
\noindent

 The proof of the proposition reduces thus in determining the value  $a_\infty$.  

\vskip5mm
\noindent 
 \textit{Proof of the Proposition:} 
At $\beta=\beta_1$, the quenched limit $f_\infty(\beta_1)$ equals  the annealed one $\overline{f}_\infty(\beta_1)=\beta_1^2/4+\log 2$, where the term $\beta_1^2/4$ comes from the mean value of the Boltzmann factor (\textit{i.e.}\ the typical behaviour and the mean behaviour coincide at this temperature).
Since for $\beta> \beta_1$, the typical and the average behaviour are no longer the same, 
we  use the standard large deviations argument in order to make the deviant behaviour at $\beta_*$ look like the typical behaviour at $\beta_1$.

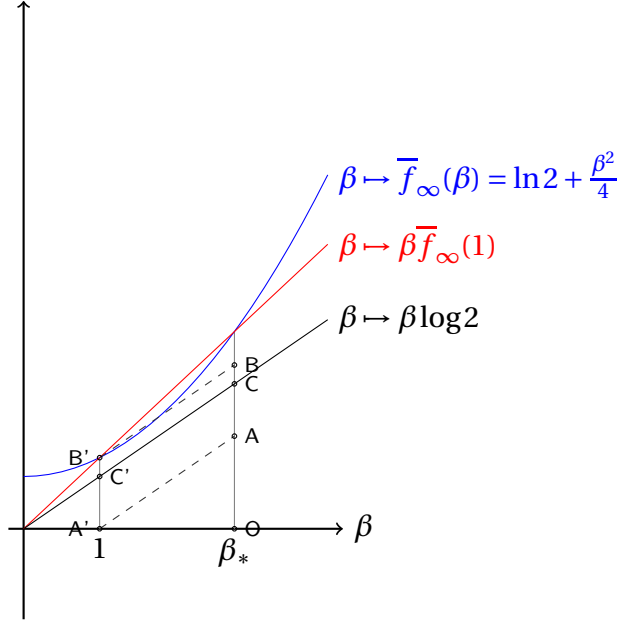
\begin{figure}[h]
\begin{center}
\begin{tikzpicture}[domain=0:4]
    \draw[->, thick,color=black] (-0.2,0) -- (4.2,0) node[right] {$\beta$};
    \draw[->, thick, color=black] (0,-1.2) -- (0,7);
   \draw[-, color=gray]    (1,0) -- (1, 0.943147) ;
   \draw[-] (1,0)--(1,0.001) node[below] {$1$};
   \draw[-] (2.772588,0)--(2.772588,0.001) node[below] {$\beta_*$};
    \draw[-, color=gray]    (2.772588,0) -- (2.772588,2.614959 ) ;
    \draw[color=blue]  plot[id=annealed] function{log(2)+0.25*(x**2)}
        node[right] {$\beta\mapsto\overline{f}_\infty(\beta) =\ln 2+ \frac{\beta^2}{4}$};
    \draw[color=red] plot[id=linear] function{x*(log(2)+0.25)} 
        node[right] {$\beta \mapsto \beta\overline{f}_\infty(1)$};
     \draw[color=black] plot[id=linearbelow] function {x*log(2)}
     node[right] {$\beta \mapsto \beta \log 2$};
     \draw[-, color =black!80,dashed] (1, 0.943147) -- (2.772588,2.171811375)  ;
     \draw[-, color =black!80,dashed] (1, 0) -- (2.772588,1.228664375) ;
          \draw (2.772588,1.228664375) circle (0.03)  node [right] {\scriptsize\textsf{A}};
         \draw (2.772588,1.9218) circle (0.03)  node [right] {\scriptsize\textsf{C}};
     \draw (2.772588,2.171811375) circle (0.03)  node [right] {\scriptsize\textsf{B}};
        \draw (2.772588,0) circle (0.03)  node [right] {\scriptsize\textsf{O}};
     \draw (1,0) circle(0.03) node [left] {\scriptsize \textsf{A'}};
     \draw (1, 0.693) circle(0.03) node [right] {\scriptsize \textsf{C'}};
     \draw (1, 0.943147) circle(0.03) node [left] {\scriptsize \textsf{B'}};
  \end{tikzpicture}
\end{center}
\caption{\label{fig:proof} 
The affine map $T$ maps $C'$ to $C$. The length of the segment $A'B'$ corresponds to the value $f_\infty(\beta_1)$ that is parallel transported to the segment $AB$.  The segment $OB$  equals  $f_\infty(\beta_*)$.  The dashed lines $A'A$ and $B'B$ are parallel to $C'C$. }
\end{figure}

The affine mapping $T$ on measures induces a transformation on the free energies reading   
 $\overline{f}_\infty(\beta_*)=\frac{\beta_*^2}{4}+\log 2=\frac{\beta_*}{\beta_1}f_\infty(\beta_1)$. It follows that  the pre-image of the term $\beta_*^2/4$ --- coming from the average of the Boltzmann factor --- (point $C$ of the figure \ref{fig:proof}),  is $\beta_1\beta_*/4=\beta_1 \log 2=\log 2$ (point $C'$); one gets the  value of the free energy  $f_\infty(\beta_1)$  if the term $\beta_1^2/4=1/4$ is  added to this pre-image. We remark that the sheer particularity of the two temperatures $\beta_1$ and $\beta_*$ is that the pre-image of $\log 2$ is $1/4$! Therefore, to obtain the quenched limit at $\beta_*$ is enough to add to the image of $\log 2$ (\textit{i.e.}\ to the segment $OC$) the value $1/4$ (segment $CB$).

One can now easily check that 
the difference of the two  limits  $f_\infty(\beta_*)$ and $f_\infty(\beta_1)$, is simply given by
the segment $OA$:
\[
f_\infty(\beta_*)-f_\infty(\beta_1)=(\beta_*-\beta_1)\log 2
.\]

Hence, 
\[f_\infty(\beta_*)=(\beta_*-\beta_1)\log 2+f_\infty(\beta_1)=\beta_* \log 2+\frac{\beta_1^2}{4}=\frac{\beta_* ^2}{4}+\frac{1}{4}=2.1718\cdots, \]
and, moreover
\begin{eqnarray*}
a_\infty& = &\overline{f}_\infty(\beta_*)-f_\infty(\beta_*)\\
& = & \frac{\beta_*}{\beta_1}f_\infty(\beta_1)-f_\infty(\beta_*)\\
& = &\frac{\beta_*\beta_1}{4}-\frac{\beta_1^2 }{4}.
 \end{eqnarray*}

One can  check that the value of $f_\infty(\beta_*)$  is slightly  
 lower than the  bound one can obtain by  making use of the spherical model    ($2.2058\cdots$).
\eproof

\vskip5mm
\noindent
\textit{Proof of the Theorem:}
For $\beta_*$, we have
\begin{eqnarray*}
s(\mu_{\beta_*}) & = &
\lim_{n\rightarrow\infty}
\frac{1}{n}S(\mu_{n,\beta_*,J})=-\lim_{n\rightarrow\infty}\frac{1}{n}\sum_\sigma\mu_{n,\beta_*}(\sigma|J)
 \log  \mu_{n,\beta_1}^{\beta_*/\beta_1}(\sigma|J)\frac{Z_n^{\beta_*/\beta_1}(\beta_1,J)}{Z_n(\beta_*,J)}\\
& = &-\lim_{n\rightarrow\infty}\frac{1}{n}\sum_\sigma\mu_{n,\beta_*}(\sigma|J)
 \log  \mu_{n,\beta_1}^{\beta_*/\beta_1}(\sigma|J)-\alpha_\infty,
\end{eqnarray*}
and, by  the positivity of the entropy one checks readily that
\[\lim_{n\rightarrow\infty}\frac{1}{n}\sum_\sigma\mu_{n,\beta_*}(\sigma|J)
 \log  \mu_{n,\beta_1}(\sigma|J)\leq -\frac{\beta_1}{\beta_*}\alpha_\infty
=\frac{1}{4\beta_*}-\frac{1}{4}.\]
In the following, we shall show that this inequality is saturated.
For this we introduce  a slightly different notation.

Let  $W_{n,\beta_1}(\sigma|J)=e^{-\beta_1H_n(\sigma,J)}/2^{n\beta_1}$  be the random weight associated with each configuration $\sigma\in \Sigma_n$.
The Gibbs  measures  $\mu_{n,\beta_1}(\sigma|J)$ and
$\mu_{n,\beta_*}(\sigma|J)$  are now given by
\[\mu_{n,\beta_1}(\sigma|J)=\frac{W_{n,\beta_1}(\sigma|J)}{\sum_\sigma W_{n,\beta_1}(\sigma|J)},\]
and,
\[\mu_{n,\beta_*}(\sigma|J)=\frac{W_{n,\beta_1}^{\beta_*/\beta_1}(\sigma|J)}{\sum_\sigma W_{n,\beta_1}^{\beta_*/\beta_1}(\sigma|J) }.\] \\

We have  indeed, from the high  temperature results,
\[
\lim_{n\rightarrow \infty}\frac{1}{n}E_J \log \sum_\sigma W_{n,\beta_1}(\sigma|J)=\lim_{n\rightarrow \infty}\frac{1}{n}\log E_J\sum_\sigma\frac{e^{-\beta_1 H_n(\sigma,J)}}{2^{n\beta_1}}=\frac{\beta^2_1}{4}+\log 2-\beta_1 \log 2=\frac{1}{4},\]
and, from the previous proposition, 
\[
\lim_{n\rightarrow \infty}\frac{1}{n}E_J \log \sum_\sigma W_{n,\beta_1}^{\beta_*/\beta_1}(\sigma|J)
=\lim_{n\rightarrow \infty}\frac{1}{n}E_J\log \sum_\sigma\left(\frac{e^{-\beta_1 H_n(\sigma,J)}}{2^{n\beta_1}}\right)^{\frac{\beta_*}{\beta_1}}=\frac{\beta_*^2}{4}-\beta_* \log 2+\frac{\beta_1^2}{4}=\frac{1}{4},
\]
{\it i.e.}\ the behaviour of the sums $\sum_\sigma W_{n,\beta_1}(\sigma|J)$ and $\sum_\sigma W_{n,\beta_1}^{\beta_*/\beta_1}
(\sigma|J)$ is the same. Thus, for the comparison of the two measures, namely for distinguishing between the behaviour of  the summands $W_{n,\beta_1}(\sigma|J)$ and $W_{n,\beta_1}^{\beta_*/\beta_1}
(\sigma|J)$ we need  additional information.

We introduce the relative entropy density  $s(\mu_{\beta_*}|\mu_{\beta_1})$ of the measure $\mu_{\beta_*}$ w.r.t.\ the measure $\mu_{\beta_1}$ which gives the extend to which the measure $\mu_{\beta_*}$ ``differs" from the measure $\mu_{\beta_1}$:
 
\[s(\mu_{\beta_*}|\mu_{\beta_1}):=\lim_{n\rightarrow \infty}\frac{1}{n}
S(\mu_{n,\beta_*}|\mu_{n,\beta_1})=\lim_{n\rightarrow \infty}\frac{1}{n}
\sum_\sigma \mu_{n,\beta_*}(\sigma|J)\log \frac{\mu_{n,\beta_*}(\sigma|J)}{\mu_{n,\beta_1}(\sigma|J)}.\]
This limit exists and it is a non-negative function vanishing 
in the case the two measures are equal. 
We notice  moreover that
\begin{eqnarray*}
s(\mu_{\beta_*}|\mu_{\beta_1})& = &\lim_{n\rightarrow \infty}\frac{1}{n}
\sum_\sigma \mu_{n,\beta_*}(\sigma|J)\log \frac{\mu_{n,\beta_*}(\sigma|J)}{\mu_{n,\beta_1}(\sigma|J)}\\
& = & \lim_{n\rightarrow \infty}\frac{1}{n}
\sum_\sigma \mu_{n,\beta_*}(\sigma|J)\log\frac{W_{n,\beta_1}^{\beta_*/\beta_1}(\sigma|J)}{W_{n,\beta_1}(\sigma|J)}\\
& = &\lim_{n\rightarrow \infty}\frac{1}{n}
\sum_\sigma \mu_{n,\beta_*}(\sigma|J)\log W_{n,\beta_1}^{{\frac{\beta_*}{\beta_1}-1}}(\sigma|J).
\end{eqnarray*}

Obviously, $W_{n,\beta_1}^{\beta_*/\beta_1}(\sigma|J)\leq\sum_\sigma W_{n,\beta_1}^{\beta_*/\beta_1}(\sigma|J)$. Hence,
\[\limsup_{n\rightarrow \infty}\frac{1}{n}\sum_\sigma \mu_{n,\beta_*}(\sigma|J)\log W_{n,\beta_1}^{{\frac{\beta_*}{\beta_1}-1}}(\sigma|J)=
\lim_{n\rightarrow \infty}\frac{1}{n}
\log \left(\sum_\sigma W_{n,\beta_1}^{\beta_*/\beta_1}(\sigma|J)\right)^{1-\frac{\beta_1}{\beta_*}}.\]
and, consequently,
\[s(\mu_{\beta_*}|\mu_{\beta_1})=\lim_{n\rightarrow \infty}\frac{1}{n}
\log \left(\sum_\sigma W_{n,\beta_1}^{\beta_*/\beta_1}(\sigma|J)\right)^{1-\frac{\beta_1}{\beta_*}}=\frac{1}{4\beta_*}(\beta_*-\beta_1)\]
where the equality of the limsup and the limit
 is a consequence of the positivity of $s(\mu_{\beta_*}|\mu_{\beta_1})$.

 Using now the functional definition of the measure one gets
\begin{eqnarray*}
s(\mu_{\beta_*}|\mu_{\beta_1})& = &\lim_{n\rightarrow \infty}\frac{1}{n}
\sum_\sigma \mu_{n,\beta_*}(\sigma|J)\log \frac{\mu_{n,\beta_*}(\sigma|J)}{\mu_{n,\beta_1}(\sigma| J)}\\
& =  & \lim_{n\rightarrow \infty}\frac{1}{n}
\sum_\sigma \mu_{n,\beta_*}(\sigma|J)\log \mu_{n,\beta_1}^{\frac{\beta_*}{\beta_1}-1}(\sigma|J)+\alpha_\infty\\
& = &
\frac{1}{4\beta_*}(\beta_*-\beta_1).
\end{eqnarray*}
Recalling that $\alpha_\infty=\frac{\beta_1}{4}(\beta_*-\beta_1)$, it follows immediately that
\[\lim_{n\rightarrow \infty}\frac{1}{n}
\sum_\sigma \mu_{n,\beta_*}(\sigma|J)\log \mu_{n,\beta_1}(\sigma|J)=
\frac{1}{4\beta_*}-\frac{1}{4}=-\frac{\beta_1}{\beta_*}\alpha_\infty\]
which proves the theorem.
\eproof

\vskip0.5mm

\noindent
{\it Remarks: } 
 Another interesting quantity is the relative entropy density $s(\mu_{\beta_*}|\nu)$ of the measure $\mu_{n,\beta_*} $
w.r.t.\ the uniform measure $\nu_n(\sigma)$ :
 \begin{eqnarray*}
s(\mu_{\beta_*}|\nu) & = &\lim_{n\rightarrow\infty}\frac{1}{n}S(\mu_{n,\beta_*}|\nu_n)
=\lim_{n\rightarrow\infty}\frac{1}{n}\sum_\sigma\mu_{n,\beta_*}(\sigma|J)\log \frac{\mu_{n,\beta_*}(\sigma|J)}{\nu_n(\sigma)}\\
& = & -s(\mu_{\beta_*})+\log 2\\
& = & \log 2.
\end{eqnarray*}
(We recall that  $s(\mu_{\beta_1}|\nu)=-s(\mu_{\beta_1})+\log 2=\frac{\beta_1^2}{4}=\frac{1}{4})$.
 
 One can also easily check that  the value of the limit  $\alpha_\infty$ corresponds to the entropy difference   $\alpha_\infty=s(\mu_{\beta_1})-s(\mu_{\beta_*})$.



\section{Concluding remarks}

In this note  we  showed that the mean entropy of the Gibbs measure vanishes at the inverse temperature $\beta_*=4 \log 2$. A related question concerns the Hausdorff dimension of the support of the Gibbs measure. From our result on the entropy 
 one can easily show that this dimension vanishes at 
 $\beta_*$.

A last observation concerns  the 
value of the temperature $\beta_*$: it is obtained from the relationship
between the  free energies $\bar{f}_\infty(\beta_*)$ and $f_\infty(1)$; 
moreover, one can readily  check that 
 $\beta_*=\beta^2_c$, where
$\beta_c=2\sqrt{\log 2}$ is the critical temperature of the Random Energy Model (REM). The REM  is  defined by $2^n$ energy 
levels $E_i (i=1,\cdots,n)$, a family of random, independent, identically
distributed random variables;  many results are qualitatively the same 
as those of the SK model.  It would be interesting to clarify   this relationship in order to obtain some information on the   behaviour  and properties of the Gibbs measure at low temperatures. 
Both  $\beta_c$ and $\beta_*$ are to be compared with
the value at $\beta_1\equiv 1$, {\it i.e.}\ the maximum value of $\beta$ where the free energies of the two models coincide. What  we learn by the comparison of the two models is that the Gibbs measure of the SK 
has seemingly a  richer structure
than for the REM.  As a matter of fact, the entropy of the REM vanishes at
$\beta_c$ while the entropy of the SK  model is still strictly positive at this point.


\begin{thebibliography}{99}
{\footnotesize
\bibitem{ALR} Aizenman, M., Lebowitz, J. L., Ruelle, D.: 
                Some Rigorous Results on the Sherrington-Kirkpatrick Spin 
                 Glass Model. 
                Commun. Math. Phys. {\bf 112}, 3--20 
               (1987).
\bibitem{ASS} Aizenman, M., Sims, R., Starr, S.,L.: An extended variational
                principle for the SK spin-glass model. Phys. Rev. B,
                 {\bf 6821}(21): 4403, (2003).
\bibitem{Der} Derrida, B., Random energy model: An exactly solvable model of disordered systems. Phys. Rev. {\bf B4}, 2613--2626 (1981).
 \bibitem{G}  Guerra, F.: Broken replica symmetry bounds in the 
              mean field spin glass model. Comm. Math Phys. {\bf 233}(1),
               1--12 (2003).
\bibitem{GT} Guerra, F., Toninelli, F.:  The thermodynamic limit in
        mean field spin glass models. 
                    Commun. Math. Phys. {\bf 230}(1), 71--79 (2002). 
\bibitem{K1} Koukiou, F.: The low temperature free energy of the Sherrington-Kirkpatrick spin glass model. 
                Eur. Lett. {\bf 33}, 95--98 (1996).
\bibitem{K2} Koukiou, F.: The ground state energy of the mean field spin glass model, ArXiv:0806.1380.
 \bibitem{P}  Parisi, G.: A sequence of approximated solutions to the 
                  Sherrington-Kirkpatrick
                    model for spin glasses. J. Phys. {\bf A 13}, L115--L121
           (1980).
\bibitem{PS}  Pastur, L., Shcherbina, M.V.: Absence of  self-averaging
                     of the order  parameter in the Sherrington-Kirkpatrick
                    model. J. Stat. Phys. {\bf 62}, 1--19 (1991).

\bibitem{SK}  Sherrington, D., Kirkpatrick, S.: Solvable model of a spin 
                glass. Phys. Rev. Lett. {\bf 35}, 1792--1796 (1975).
                Infinite-ranged  models of spin-glasses. Phys.Rev. {\bf 17},
                4384--4403 (1978).
 \bibitem{T1} Talagrand, M.: The Parisi formula. Ann. of   Mathematics {\bf 163}, 221--263 (2003).
}
 \end{thebibliography}
\end{document}